\documentclass[journal]{IEEEtran}
\usepackage{multirow}
\usepackage{array}
\usepackage{tabu}
\usepackage{array,booktabs,arydshln,xcolor}
\usepackage{tikz}
\usetikzlibrary{matrix,shapes,arrows,positioning,chains}
\usepackage{amsmath}
 
\newcommand\norm[1]{\left\lVert#1\right\rVert}
\usepackage{amssymb}
\usepackage{enumerate}
\usepackage{algorithm}
\usepackage{amssymb}
\usepackage{textcomp}
\usepackage{commath}
\usepackage{graphicx}
\usepackage{array}
\usepackage{tabu}
\usepackage{array,booktabs,arydshln,xcolor}
\usepackage{tikz}
\usetikzlibrary{matrix,shapes,arrows,positioning,chains}
\usepackage[noend]{algpseudocode}
\usepackage{pifont}
\usepackage{csquotes}
\usepackage{graphicx}
\usepackage{array}
\usepackage{dblfloatfix}
\usepackage{pgfplots}
\usepackage{tikzscale}

\usepackage{csquotes}
\usepackage{float}
\usepackage{cite}
\usepackage{breqn}
\usepackage{blox}
\usetikzlibrary{shapes,arrows}

\ifCLASSINFOpdf  
\else
\fi
\begin{document}
\title{Adaptive Control of  Unknown Pure Feedback  Systems with Pure State  Constraints
\thanks{The authors are with the Department of Electrical Engineering,
Indian Institute of Technology Kanpur, Kanpur 208016, India (e-mail:
pmpankajjec@gmail.com; nishchal.iitk@gmail.com.}}

\author{Pankaj Kumar Mishra
        and Nishchal K Verma}% <-this % stops a space

\maketitle

\begin{abstract}
 This paper deals with the tracking control problem for a class of unknown pure feedback system with pure state constraints on the state variables and unknown time-varying bounded disturbances. An adaptive controller is presented for such systems for the very first time. The controller is designed using the backstepping method. While designing it, Barrier Lyapunov Functions is used so that the state variables do not contravene its constraints. In order to cope with the unknown dynamics of the system, an online approximator is designed using a neural network with a novel adaptive law for its weight update. In the stability analysis of the system,  the time derivative of Lyapunov function involves known virtual control coefficient with unknown direction and to deal with such problem Nussbaum gain is used to design the control law. Furthermore, to make the controller robust and computationally inexpensive, a novel disturbance observer is designed to estimate the disturbance along with neural network approximation error and the time derivative of virtual control input. The effectiveness of the proposed approach is demonstrated through a simulation study on the third-order nonlinear system.
\end{abstract}


\IEEEpeerreviewmaketitle

\section{Introduction}
\IEEEPARstart{I}{n} recent years, the stability of the constrained nonlinear system has attracted much attention in the nonlinear control theory community. The reason behind this is its application in the industrial systems. In real-time systems, constraints can appear in different forms such as performance specification, safety, physical stoppage and saturation, and it is ineludible while designing the controller.  Dynamically it can appear as a symmetric or asymmetric bound on states, output and control input of the system.\par
The traditional controller design for nonlinear unconstrained system lack practicability. In \cite{TEE2009918}, Tee et al.  have proposed a  controller for nonlinear systems with constant output constraints. To prevent constraints, authors have proposed Barrier Lyapunov Function (BLF), which approach infinity when its argument approach certain limit. In \cite{TEE20112511},  Tee et al.  have proposed a BLF based controller for nonlinear systems with time-varying output constraints. With the above pioneering  works, researchers have started paying attention in the field of controller design for nonlinear system with constraints. A lot of significant related works have been received in recent years. In \cite{5499019},  Ren et al. have studied a  BLF based adaptive controller for a nonlinear system with time-varying constraints on the output. In \cite{6651788,4689317, 7031439}, BLF based controller has been studied to tackle practical output constraint for electrostatic micro-actuators, flexible crane system and a wind turbine system.\par With the progress in control of a nonlinear constrained system, in \cite{tanda1}   Tee et al. have studied controller design for a nonlinear system with partial state constraint. 
% There are very few works  \cite{7429795, LIU201670, LIU2017143, 7571100,7435341, 7786843} devoted on controller design for nonlinear system with full state constraints. 
Liu et al. in \cite{7429795} and \cite{LIU201670} have studied a BLF based adaptive backstepping control for strict feedback and pure-feedback single input single output (SISO) nonlinear system having a static constraint on all the states, respectively. In \cite{LIU2017143}, the authors have studied control of SISO nonlinear system having unknown control gain and static constraint on all the states. In \cite{7571100, 7435341}, the authors have studied BLF based adaptive control of multi input multi output (MIMO) nonlinear system having static and symmetric constraints on all the states. In \cite{7786843}, the authors have used novel BLF for control of time-varying state constrained SISO nonlinear system.  In \cite{8759967, 8815865}, the authors have studied BLF for MIMO nonlinear systems having time-varying state constraints.  \par 
Other than BLF based methodology, numerous efforts have been made, such as error transformation and model predictive control (MPC), by academia and industries to design a controller for the constrained system. In error transformation \cite{4639441, BECHLIOULIS2009532, WANG20102082, 5772916, BECHLIOULIS20141217}, there is likelihood that use of tangent hyperbolic in prescribed function ends with singularity problem and also under specific conditions of prescribed function, inordinate control input can transgress the prescribed control performance, which may lead to instability.  In MPC, constraint is accommodated in control design for linear and nonlinear system within an optimization framework by solving a finite horizon open-loop optimal control problem \cite{MAYNE2000789}. Most of the optimal control and MPC need knowledge of the dynamics of plant and are numerical and  thus depend mostly on computationally intensive algorithms for solving a control problem \cite{kirk_2016}.\par As compared to error transformation and  MPC, BLF has been extensively investigated for the controller design of   constrained system because of its ease in handling unknown system dynamics, uncertainties and disturbances by integrating robust adaptive backstepping or sliding mode control methodology.  So far, the literature related to BLF based controller design has been limited for the system having a static and time-varying constraint on all the states \cite{8759967,8418495,CAO2019108608, ptv}.
% \cite{LIU201670,8668697,8952870,LI2018444, 8353903, 8290541,8879661,pfc1, pfc2, 8880612, 8544037, 8603838, 8896007, 8425067, 8732686, XI2019108,SONG2018314,8948553, 8603838,8815865,8759967,8418495,CAO2019108608, ptv}
However, there are still a lot of challenging problems which are yet to be explored by the researchers working in this field for the constrained system. One such problem is to control a  system with pure state constraints on the state variable. Pure state constraints are state variable inequality constraints (SVICs), which are expressed in terms of time and the state variables. Such constraints frequently arise in the area of management science and economics, mechanics, and aerospace engineering \cite{doi:10.1137/1037043}.\par
The aforementioned problem acts as motivation for this paper. As a solution,   we design a  BLF  based robust adaptive backstepping control law using a neural network (NN) for an unknown nonlinear pure feedback system with pure state constraints and time-varying bounded disturbances. The following are essential steps which outline the design of the controller in this paper:\par
Similar to traditional backstepping approach, the first step involves the transformation of state variables to error variables using the virtual control input. Second, the construction of error variable inequality constraints (EVICs) for error variables using the SVICs. Third,  construction of BLF using EVICs and the calculation of its time derivative. The time derivative of BLF is calculated in an early stage to avoid the unnecessary steps of calculating similar time derivative of BLF in the controller design and stability analysis. The time derivative of BLF involves a virtual control coefficient with unknown direction. To deal with such problem,  Nussbaum gains \cite{1220768, 5587877, 5723705}  is used in the proposed methodology. The fourth step includes the design of NN for the approximation of unknown function involving uncertain dynamics of the system. The fifth step consists of the design of a disturbance observer for the estimation of disturbances along with the derivative of virtual control input and NN approximation error. Finally, using the stability analysis  Nussbaum gain based backstepping control law,  and an adaptive law for the weight updates of NN is developed. In brief, our contributions  are as follows:
\begin{enumerate}
    \item As mentioned in Table \ref{tab1}, compared to state-of-the-art problems, where the controller is designed for the system with static and time-varying state constraint, here a novel controller is proposed for the system with pure state constraints. Moreover, the system is considered as a pure feedback SISO nonlinear system.
    \item A BLF based disturbance observer is proposed to estimate the expression involving derivative of virtual control input, external disturbance and NN approximation error.  This makes the controller robust and computationally efficient.
\end{enumerate}

 \begin{table}[h]
 \caption{Comparison with state-of-the-art problems }
 \scalebox{0.7}{
\begin{tabular}{cccccc}
\hline
& \multicolumn{3}{c}{Types of  State Constraint}& \multicolumn{2}{c}{Structure of Nonlinear Systems} \\ \hline
 & \multicolumn{1}{c}{Constant} & \multicolumn{1}{c}{\begin{tabular}[c]{@{}c@{}}Time-varying \\$\Psi_i( t)$\end{tabular}} & \multicolumn{1}{c}{\begin{tabular}[c]{@{}c@{}}Pure State \\Constraint \\$\Psi_i( \overline \xi_i, t)$\end{tabular}} & \multicolumn{1}{c}{\begin{tabular}[c]{@{}c@{}}Strict-\\ feedback\end{tabular}} & \multicolumn{1}{c}{\begin{tabular}[c]{@{}c@{}}Pure-\\ feedback\end{tabular}}  \\ \hline
\cite{8668697,8952870,LI2018444, 8353903, 8290541,8879661} & \checkmark & $\times$  &$\times$  &\checkmark & $\times$ \\
\cite{LIU201670,pfc1, pfc2} & \checkmark & $\times$  &$\times$  &$\times$ & \checkmark\\

\cite{8896007, 8425067, 8732686, XI2019108,SONG2018314,8948553, 8603838,8815865,8759967} & $\times$ &\checkmark  &$\times$  &\checkmark & $\times$\\
\cite{8418495,CAO2019108608, ptv} & $\times$ &\checkmark  &$\times$  &$\times$ &\checkmark \\
Proposed  problem & $\times$ & $\times$  &\checkmark  &$\times$ & \checkmark\\\hline
\end{tabular}
\label{tab1}
}
\end{table}
The paper is organized as follows. In Section II, we present the system description and control objectives.  This section also presents some assumptions, definition and Lemmas for the stability analysis of the system. Section  III consists of five subsections whose first subsection discusses the construction of EVICs for error variables. The second subsection discusses the construction of BLF and calculation of its time derivative. The third subsection discusses the construction of NN for the approximation of unknown term involved in the time derivative of BLF derived in the second subsection; The fourth subsection discusses the construction of disturbance observer for the robustness of the system, and the fifth subsection discusses the steps to design an adaptive  controller using the decoupled backstepping technique. Section IV discusses the theorem for the boundedness of all the signals in the closed-loop of the system. Section V discusses the effectiveness of the proposed methodology using the simulation examples. Finally, Section VI concludes the paper.\\
Following are  some basic notations which will be used throughout the paper:
\begin{itemize}
\item $\mathbb{N}_m := \{1, \ldots, m\}$.
\item $arctan (x)= \tan ^{-1} x$.
% \item The symbol \lq$\sup$\rq ~ denotes supremum. The supremum of a set is its least upper bound.
% \item The symbol \lq$\inf$\rq ~ denotes infimum. The infimum of a set is its greater lower bound. 
\end{itemize}
\section{System description and problem statement}
Consider the following  SISO nonlinear pure-feedback system 
\begin{equation}\label{sys}
\begin{split}
\dot x_i&=f_i\left(\bar{x}_i,x_{i+1}\right)+\beta_ix_{i+1}+d_i\left(\bar{x}_i,t\right)\\
\dot x_n&=f_n\left(\bar{x}_n,u\right)+\beta_n u+d_n\left(\bar{x}_n,t\right)\\
y&=x_1
\end{split}
\end{equation}
where $x_i\in \mathbb{R}$,  $\forall i \in \mathbb{N}_n, y\in \mathbb{R}$ and $u\in \mathbb{R}$ are the $ith$ state,  the output, and the control input of the  system, respectively; $\overline x_i=[x_1, \hdots, x_i]^T \in \mathbb{R}^i$; $f_i\left(\bar{x}_i,x_{i+1}\right) \in \mathbb{R} , \forall i \in \mathbb{N}_n$ are smooth  unknown nonlinear function; $\beta_i\in \mathbb{R}$ and  $d_i \in \mathbb{R} , \forall i \in \mathbb{N}_n$ are constant control coefficient and unknown time-varying bounded disturbance, respectively. For simplicity of presentation, denote $x_{n+1}=u$. 
% The structure of (\ref{sys}) is shown in Fig. \ref{sysfig}.
In this study states are considered to be constrained such that, $\abs{x_i}<\Psi_i(\overline x_i, t)$, where $\Psi_i \in \mathbb{R}$ is a known nonlinear SVIC on the state variable.\\
% \vspace{-0.7cm}
% \begin{figure}[H]
%     \centering
%     \includegraphics[width=7cm,height=5cm,keepaspectratio]{plantppp.png}
%     \caption{ Nonlinear pure feedback system}
%     \label{sysfig}
% \end{figure}\par
{\textit{Problem Statement:}} The control objective of the paper is to design a NN based adaptive controller for  (\ref{sys}) such that (i) output $y$ tracks the desired output $y_d$; (ii) all the closed-loop signals are guaranteed to be bounded;  and (iii) all the system states do not contravene there SVICs.\\
Following are the assumptions which will be needed to achieve the control objective.\\
% \textit{Assumption 1:} The state variables, $x_i$, $\forall i \in \mathbb{N}_n$  are observable.\\
\textit{Assumption 1 \cite{LIU2017143}:}  The control coefficient $\beta_i\neq 0$, $\forall i \in \mathbb{N}_n$.\\
\textit{Assumption 2 \cite{1220768}:} The unknown time-varying $\text{disturbance}$ $d_{i}(\overline x_i,t)$ is bounded and there exist some positive constant $d_0$ such that $\abs {\dot d_{i}(\overline x_i,t)} \le d_0$ $\forall i \in \mathbb{N}_n$.\\
{\textit{Assumption 3}} \textit{\cite{7927471}:} {\textit{If $x_i \in  L_\infty$ then the time derivative of $ \frac{\partial \psi}{\partial x_1}, \hdots, \frac{\partial \psi}{\partial x_i}$ exist and it is bounded.}
% \textit{Assumption 3:} The unknown time-varying disturbances $d_{i}(\overline x_i,t)$ is bounded and there exist some positive constant $d_0$ such that $\norm {d_{i}(\overline x_i,t)} \le d_o$ $\forall i \in \mathbb{N}_n$.\\
\textit{Definition 1  \cite{Nussbaum1983}} : The  function $\mathcal{N}(\zeta)$ is said to be Nussbaum, if it holds the following property:\\
\begin{equation}
\begin{split}
\underset{s \rightarrow \infty}{\lim} \sup \frac{1}{s}\int_{0}^s \mathcal{N}(\zeta)d\zeta = +\infty\\
\underset{s \rightarrow \infty}{\lim} \inf \frac{1}{s}\int_{0}^{s} \mathcal{N}(\zeta)d\zeta = -\infty.
\end{split}
\end{equation}
There are many functions which can be considered as a Nussbaum function such as $e^{\zeta^2}\cos((\pi/2) \zeta)$ and $\zeta^2cos(\zeta)$. In this  paper, we have used $\mathcal{N}(\zeta)=\zeta^2cos(\zeta)$  as a Nussbaum function. \\
Following are the Lemmas which will be used throughout the paper\\
\textit{Lemma 1 \cite{1220768}:} Let $V(t)\ge 0$ and $\zeta(t)$ be  smooth functions defined on $[0, t_f)$ and  $\mathcal{N}(\zeta(t))$  be an even smooth Nussbaum function. If the following inequality holds: \\
\begin{align}
    V(t)\le \kappa_1+e^{-\kappa_2t}\int_{0}^{t}\left({\beta_0\mathcal{N}(\zeta)}+1\right)\dot\zeta e^{\kappa_2 \tau}d\tau
\end{align}
where $\kappa_1$ and $\kappa_2$ are positive constant, and $\beta_0$ is a non-zero constant, then $V(t)$, $\zeta(t)$ and $\int_{0}^{t}{\beta_0\mathcal{N}(\zeta)}\dot\zeta d\tau$ are bounded on $[0, t_f)$.\\
% \textit{Lemma 2 \cite{8759967} :} For a bounded function $\mathfrak{f}( v)=\frac{2A}{\pi}\arctan \left(\frac{\pi  v}{2A}\right)$ with bound $A$ i.e. $\abs{\mathfrak{f}(v)}\le A$, $\exists$  $\left( v_1 \in \mathbb{R} ~\text{and}~  v_2 \in \mathbb{R}\right)$ such that the function  $\mathfrak{f}(v)$ with  $ v= v_1 + v_2$ can be written as $\mathfrak{f}( v)= v_1+\mathfrak{g}( v_1, v_2)$ where $\mathfrak{g}( v_1, v_2)= v_2-\frac{p^2 v^2}{1+p^2 v^2} v$, 
% $p=\frac{\pi \bar\mu}{2A}$, and $0<\bar \mu<1$ .\\
% \textit{Corollary 1 \cite{8759967}:} For any $\mathfrak{h}\in\mathbb{R}$, $\exists$ $ v_2 \in \mathbb{R}$ such that, if $v_2=\mathfrak{h}(\sigma_1+\rho(\sigma_2-\sigma_1))$, where 
% \begin{align}\label{root}
% \sigma_1&=\frac{(1-2p^2 v_1^2-\sqrt[]{1-4p^2v_1^2})}{2p^2 v_1}, ~\text{and} \\
% \sigma_2&=\frac{(1-2p^2 v_1^2+\sqrt[]{1-4p^2v_1^2})}{2p^2 v_1}.
% \end{align}
% then $\mathfrak{h}\mathfrak{f}(v)<\mathfrak{h} v_1$.\\ 
\textit{Lemma 2 \cite{8759967}:} For $\vartheta=\mathfrak{f}( v)=\frac{2A}{\pi}\arctan \left(\frac{\pi  v}{2A}\right)$, $\exists$  ($ v_{1} \in \mathbb{R}$, $ v_{2} \in \mathbb{R}$) such that for any $\mathfrak{h}\in\mathbb{R}$, if  $ v= v_{1} + v_{2}$ and $v_{2}=\mathfrak{h}(\sigma_{1}+\rho(\sigma_{2}-\sigma_{1}))$ where $0<\rho<1$,
\begin{equation}\label{root}
\begin{split}
\sigma_{1}&=\frac{(1-2p^2 v_{1}^2-\sqrt[]{1-4p^2v_{1}^2})}{2p^2 v_{1}},\quad  \\
\sigma_{2}&=\frac{(1-2p^2 v_{1}^2+\sqrt[]{1-4p^2v_{1}^2})}{2p^2 v_{1}}
\end{split}
\end{equation}
and $p=\frac{\pi \bar\mu}{2A}$, $0<\bar \mu<1$, then  we have $\mathfrak{h}\vartheta\le\mathfrak{h} v_{1}$.\\  
% \textit{Lemma 3:} If $f(x)>g(x)$ for $a<x<b$ then $\int_a^bf(x)dx>\int_a^bg(x)dx$.\\
\textit{Lemma 3 \cite{5499019}:} For any $\psi \in \mathbb{R}$,  $\log\frac{\psi^2}{\psi^2-z^2}<\frac{z^2}{\psi^2-z^2}$ in the interval $\abs{z} < \abs{\psi}$.\\
% \textit{Lemma 4:} If $f(x)>g(x)$ for $a<x<b$ then $\int_a^bf(x)dx>\int_a^bg(x)dx$.\\
% \textit{Lemma 5:} If the matrix A, and B are of same size, then $\norm{A\circ B}_1=\Tr(AB^T)=\Tr(BA^T)$.\\
% \textit{{Lemma 6 \cite{580878}:}} If the signal $e(t)\in L_2 \cap L_{\infty}$  and $\dot e(t) \in L_{\infty}$, then $\underset{t \rightarrow \infty}{\lim}e(t)=0$.\\
% \textit{Lemma 4 \cite{Ryan1991}:} For any given positive constant $t_f>0$, if the solution of the resulting closed-loop system is bounded on the interval $[0,t_f)$, then $t_f=\infty$ .\\
\section{Robust adaptive backstepping controller}
Following the   traditional approach of designing controller using backstepping, let  us define few variables which will be used in controller design. Let
\begin{align} \label{err}
z_i&=x_i-\vartheta_{i-1}~~~ \forall i \in \mathbb{N}_n
\end{align}
where $\vartheta _0$ is the desired output, i.e. $\vartheta _0=y_d$; $z_i \in \mathbb{R} $ and $\vartheta_{i-1} \in \mathbb{R} $  is an error variable and a virtual control input, respectively $\forall i \in \mathbb{N}_n$.    Let, $\psi_i(\overline x_i, t)$ be the  error variable inequality constraint (EVIC)  for  the error variable $z_i$. The objective is to design $\psi_i(\overline x_i, t)$  such that, if the error variable follows EVIC, then the corresponding state variable must also follow their SVIC. In other words, $\psi_i(\overline x_i, t)$ must be designed such that, the  condition below
\begin{align}\label{cnd1}
 \abs{z_i}< \psi_i(\overline x_i, t) \implies \abs{x_i}<\Psi_i(\overline x_i, t)   
\end{align}
 holds true $\forall i \in \mathbb{N}_n$. In the next  subsection we will discuss the design of EVICs for the error variables.
\subsection{Designing EVICs  for the error variables}
In order to design EVICs, the bound on virtual control input $\vartheta_{i-1}$ must be known beforehand. In this paper, we have designed virtual control input as
 \begin{align}\label{nvi}
     \vartheta_{i}=\frac{2A_{i}}{\pi}\arctan \left(\frac{\pi  v_{i}}{2A_{i}}\right),
 \end{align}
 where $v_{i}$ is a new virtual control input corresponding to $ \vartheta_{i}$, and $A_i$ is bound on $\vartheta_i$. \\
 \textit{Note:} From (\ref{nvi}) it is obvious that $\vartheta_i$ will not go beyond $A_i$.\\
 As the condition $\abs{\vartheta_i}<A_{i}$ is fulfilled, hence  using (\ref{err}) we can write
\begin{align}\label{cnd2}
    \abs{x_i}<\abs{z_i}+A_{i-1}.
\end{align}
Let $\psi_i(\overline x_i, t)$ be designed using the following relation 
 \begin{align}
\Psi_i(\overline x_i, t)=\psi_i(\overline x_i, t)+A_{i-1}.\label{con}     
 \end{align}
 then using (\ref{cnd2}) and (\ref{con}), we can say that the condition (\ref{cnd1}) will always hold true for $\abs{z_i}< \psi_i(\overline x_i, t)$, i.e. the state will not contravene its constraint or  $\abs{x_i}<\Psi_i(\overline x_i, t)$ is guaranteed. In the next subsection, we will see  the design of Lyapunov function using  EVIC  $\psi_i(\overline x_i, t)$, and will compute the time derivative of Lyapunov function. 
 \vspace{-0.5cm}
 \subsection{Barrier Lyapunov Functions using EVICs}
 This section is dedicated to the design of  BLF and calculation of its time derivative, which will be further used for the controller design. Let $L_i$ be the BLF which is designed as 
 \begin{align}\label{L}
     L_i=\frac{1}{2}\log\frac{\psi_i^2}{\psi_i^2-z_i^2}.
 \end{align}
 \textit{Note:} The time derivative of BLF is computed to derive general expression, which will help in eliminating redundant expression with changed lower indices, during the controller design.\\
On differentiating (\ref{L}) with respect to time, we have
%  \textit{Note:}The time derivative of BLF is computed to eliminate the few steps of calculation during the backstepping controller design, which have redundant expression with changed lower indices.\\
% On differentiating (\ref{L}) with respect to time we have
\begin{align}\label{ldot}
    \dot L_i=Q_i\left(\dot z_i-\frac{z_i}{\psi_i}\dot\psi_i\right),
\end{align}
\begin{align}\label{Qi}
  \text{where~} Q_i=\frac{z_i}{\psi_i^2-z_i^2} 
\end{align}
On using (\ref{err}) in  (\ref{ldot}), we have
\begin{align}\label{ldot1}
\dot L_i=Q_i\left(\dot x_i-\dot \vartheta_{i-1}-\frac{z_i}{\psi_i}\dot\psi_i\right).
\end{align}
On using (\ref{sys}) in (\ref{ldot1})
\begin{align}\label{ldot2}
\dot L_i=Q_i\left(f_i\left(\bar{x}_i,x_{i+1}\right)+\beta_ix_{i+1}+d_i-\dot \vartheta_{i-1}-\frac{z_i}{\psi_i}\dot\psi_i\right).
\end{align}Since, the EVIC $\psi_i$ is a function of $\bar x_i$ and $t$. Therefore,  $\dot \psi_i$ can be written as
\begin{align}\label{psid}
    \dot \psi_i=\frac{\partial \psi_i}{\partial x_1}\dot x_1+\cdots+\frac{\partial \psi_i}{\partial x_i}\dot x_i+\frac{\partial \psi_i}{\partial t}.
\end{align}
Now using (\ref{sys}) in  (\ref{psid}), we have
\begin{align}
    \dot \psi_i&=\frac{\partial \psi_i}{\partial x_1}\left(f_1(\bar{x}_1,x_2)+\beta_1x_{2}+d_1\right)+\cdots \notag\\ 
    &+\frac{\partial \psi_i}{\partial x_i}\left(f_i\left(\bar{x}_i,x_{i+1}\right)+\beta_ix_{i+1}+d_i\right)+\frac{\partial \psi_i}{\partial t} \label{psid1}.
\end{align}
 Substituting  (\ref{psid1}) into (\ref{ldot2}) yields
\begin{align}\label{ldot3}
    \dot L_i&=Q_i\left(f_i\left(\bar{x}_i,x_{i+1}\right)+\beta_ix_{i+1}-\frac{z_i}{\psi_i}\left( \frac{\partial \psi_i}{\partial x_1}\left(f_1\left(\bar{x}_1,x_2\right)+\beta_1 x_2\right) \right. \right. \notag\\
    & \quad +\cdots \left. \left.+\frac{\partial \psi_i}{\partial x_i}\left(f_i\left(\bar{x}_i,x_{i+1}\right)+\beta_ix_{i+1}\right)+\frac{\partial \psi_i}{\partial t} \right) \right) \notag\\
    & \quad +Q_i\left(d_i-\frac{z_i}{\psi_i}\left(\frac{\partial \psi_i}{\partial x_1}d_1+\cdots+\frac{\partial \psi_i}{\partial x_i}d_i\right)\right)-Q_i\dot\vartheta_{i-1}.
\end{align}  
Equation (\ref{ldot3}) is further simplified as
% \begin{align}
%   \dot L_i&= Q_i\left(f_i\left(\bar{x}_i,x_{i+1}\right)-\frac{z_i}{\psi_i}\left( \frac{\partial \psi_i}{\partial x_1}\left(f_i\left(\bar{x}_1,x_2\right)+x_2\right)+\right.\right.\notag\\
%   &\quad\left.\left.\cdots+\frac{\partial \psi_i}{\partial x_{i-1}}\left(f_{i-1}\left(\bar{x}_{i-1},x_{i}\right)+x_{i}\right)+\frac{\partial \psi_i}{\partial x_i}f_{i}\left(\bar{x}_{i},x_{i+1}\right)\right)\right)\notag\\
%   &+Q_i\left(1-\frac{\partial \psi_i}{\partial x_i}\right)x_{i+1}-Q_i\frac{z_i}{\psi_i}\frac{\partial \psi_i}{\partial t}\notag\\
%   &+Q_i\left(d_i-\frac{z_i}{\psi_i}\left(\frac{\partial \psi_i}{\partial x_1}d_1+\cdots+\frac{\partial \psi_i}{\partial x_i}d_i\right)\right)-Q_i\dot\vartheta_{i-1}
% \end{align}
\begin{align}\label{ldot4}
   \dot L_i&= \xi_i+\beta_ix_{i+1}-Q_i\frac{z_i}{\psi_i}\frac{\partial \psi_i}{\partial t}\notag\\
  &+Q_i\left(d_i-\frac{z_i}{\psi_i}\left(\frac{\partial \psi_i}{\partial x_1}d_1+\cdots+\frac{\partial \psi_i}{\partial x_i}d_i\right)\right)-Q_i\dot\vartheta_{i-1},
\end{align}
where $\xi_i$ is given as
% \begin{align}\label{betai}
%     \beta_i=Q_i\left(1-\frac{\partial \psi_i}{\partial x_i}\right) ~~~\text{and}
% \end{align}
% \begin{align}\label{xi}
%     \xi_i&=Q_i\left(f_i\left(\bar{x}_i,x_{i+1}\right)+\beta_ix_{i+1}-\frac{z_i}{\psi_i}\left( \frac{\partial \psi_i}{\partial x_1}\left(f_1\left(\bar{x}_1,x_2\right)+\beta_1x_2\right) \right. \right. \notag\\
%     & \quad +\cdots \left. \left.+\frac{\partial \psi_i}{\partial x_i}\left(f_i\left(\bar{x}_i,x_{i+1}\right)+\beta_ix_{i+1}\right)+\frac{\partial \psi_i}{\partial t} \right) \right)-\beta_ix_{i+1} .
% \end{align}
\begin{IEEEeqnarray}{lll}
&& \xi_i=Q_i\left(f_i\left(\bar{x}_i,x_{i+1}\right)+\beta_ix_{i+1}-\frac{z_i}{\psi_i}\left( \frac{\partial \psi_i}{\partial x_1}\left(f_1\left(\bar{x}_1,x_2\right)+\beta_1x_2\right) \right. \right. \notag\\
&& +\cdots \left. \left.+\frac{\partial \psi_i}{\partial x_i}\left(f_i\left(\bar{x}_i,x_{i+1}\right)+\beta_ix_{i+1}\right)+\frac{\partial \psi_i}{\partial t} \right) \right)-\beta_ix_{i+1} .
\IEEEeqnarraynumspace \label{xi}
\end{IEEEeqnarray}
From (\ref{xi}), it can be seen that $\xi_i$ consists of unknown nonlinear functions  $f_1\left(\bar{x}_1,x_2\right),\hdots, f_i\left(\bar{x}_i,x_{i+1}\right)$. Hence, a learning methodology must be developed for the approximation unknown function $\xi_i$. The following subsection discusses the approximation of $\xi_i$ using NN.

\subsection{Approximation of unknown function}
As is well known Radial Basis Function (RBF) NN has universal approximation property. So, in this paper RBF NN is used as an approximation tool. To approximate $n$ unknown functions, we need $n$ NN. The RBF NN used here  has $l$ number of hidden neurons and $1$ output. The output of $ith$ NN network is given by
\begin{align}
    h_{NN}(W_i,\bar z_i)=W_i^T\phi_i(\bar z_i), ~   \bar z_i \subset \Omega_{\bar z_i}  \forall i \in \mathbb{N}_n,
\end{align}
where $\bar z_i= [\bar x_i, x_{i+1}]^T \in \mathbb{R}^{i+1}$ is the input vector,  $W_i \in  \mathbb{R}^{l}$ is the weight vector, and $\phi_i(\bar z_i) \in \mathbb{R}^l$ is a basis vector of RBF NN defined on a compact set $\Omega_{\bar z_i}$, such that
\begin{align}
    \phi_i(\bar z_i)=\exp\left(\frac{{-\norm{\bar z_i-c_i}}^2}{b_i} \right),~~ \forall i \in \mathbb{N}_n,
\end{align}
where $c_i \in \Omega_{\bar z_i}$ is the centre of the receptive field and $b_i \in \mathbb{R}$ is the width of  Gaussian function. From the definition of $\phi_i(\bar z_i)$, we find that it is bounded. Let say $\bar \phi$ be the upper bound of $\phi_i(\bar z_i)$ then
\begin{align}\label{bnd1}
    \norm{\phi_i(\bar z_i)}\le\bar \phi ~~~~ \forall i\in \mathbb{N}_n.
\end{align}
Let there exists an ideal weight vector $W_i^* \in \mathbb{R}^l  $ such that  $ \forall i\in \mathbb{N}_n$ 
\begin{align}\label{w}
        \xi_i=W_i^{*T}\phi_i(\bar z_i)+\epsilon_i(\bar z_i),
\end{align}
where $\epsilon_i \in \mathbb{R}$ is the approximation error which is bounded, i.e.  $\norm{\epsilon}\le \bar \epsilon $ with $\bar \epsilon $ being an unknown positive constant\\
The ideal weight $W_i^*$ is defined as follows
\begin{align}
    W^*_i=\arg \underset { W_i^* \in \mathbb{R}^{l }}{\min}\left\lbrace \underset {\bar z_i \in \Omega _ {z_i}} {\sup} \left( h_{NN}(W_i^*,\bar z_i)-  h_{NN}(W_i,\bar z_{i})\right)\right\rbrace.
\end{align}
Since the ideal weight vector is unknown, so it must be estimated. Let $\hat W_i$ be an estimation of ideal weight  vector $W_i^*$ such that $ \forall i\in \mathbb{N}_n$
\begin{align}
     \hat \xi_i=\hat W_i^{T}\phi_i(\bar z_i),
\end{align}
where $\hat \xi_i$ is the estimation of an unknown nonlinear function $\xi_i$.\\
\textit{Remark 1:} Filtered $x_{i+1}$ is used in the RBF input vector $\bar z_i= [\bar x_i, x_{i+1}]^T \in \mathbb{R}^{i+1}$  to circumvent algebraic loop problems \cite{4601100}. $\hfill \square$\\
On substituting (\ref{w}) in (\ref{ldot4}), we have 
\begin{align}\label{ldot5}
   \dot L_i&= W_i^{*T}\phi_i(\bar z_i)+\epsilon_i(\bar z_i)+\beta_ix_{i+1}-Q_i\frac{z_i}{\psi_i}\frac{\partial \psi_i}{\partial t}\notag\\
  &+Q_i\left(d_i-\frac{z_i}{\psi_i}\left(\frac{\partial \psi_i}{\partial x_1}d_1+\cdots+\frac{\partial \psi_i}{\partial x_i}d_i\right)\right)-Q_i\dot\vartheta_{i-1}.
\end{align}
The last term of expression (\ref{ldot5}) involves the derivative of virtual control input. It is well known that in backstepping based controller design, the derivative of virtual control explodes to a big expression and increases the computation complexity of the controller. Moreover, the second and fifth term of  (\ref{ldot5}) involves unknown bounded approximation error and disturbance.  A new variable $\varepsilon_i$  
\begin{align}\label{var}
    \varepsilon_i&=Q_i\left(d_i-\frac{z_i}{\psi_i}\left(\frac{\partial \psi_i}{\partial x_1}d_1+\cdots+\frac{\partial \psi_i}{\partial x_i}d_i\right)\right)\notag\\
    &\quad +\epsilon_i(\bar z_i)-Q_i\dot\vartheta_{i-1},
\end{align}
is introduced as an unknown, uncertain term. Rewriting (\ref{ldot5}) using the variable defined in (\ref{var}), we have
\begin{align}\label{ldot7}
    \dot L_i&= W_i^{*T}\phi_i(\bar z_i)+\beta_ix_{i+1}-Q_i\frac{z_i}{\psi_i}\frac{\partial \psi_i}{\partial t}+\varepsilon_i.
\end{align}
% Using (\ref{err}), we can rewrite (\ref{ldot6}) as
% \begin{align}\label{ldot7}
%     \dot L_i&= Q_iW_i^{*T}\phi_i(\bar z_i)+\beta_iz_{i+1}+\beta_i\vartheta_{i}-Q_i\frac{z_i}{\psi_i}\frac{\partial \psi_i}{\partial t}+\varepsilon_i.
% \end{align}
To make the  controller robust and computationally efficient,  the unknown variable $\varepsilon _i$ must be estimated. A novel disturbance observer is designed in next subsection for the estimation of $\varepsilon_i$.
\vspace{-0.5cm}
\subsection{Disturbance observer design using BLF}
 Let $\varepsilon _i$ be the observer variable. To estimate   $\varepsilon _i$, an auxiliary variable is introduced. It is defined as 
\begin{align}\label{delta}
    \delta_i=\varepsilon_i-k_{\varepsilon_i}L_i,~~~~~ \forall i \in \mathbb{N}_n.
\end{align}
% \textit{Assumption 4 \cite{8759967}:} The observer variable $\varepsilon_i$ to be estimated, is bounded and $\forall i \in \mathbb{N}_n$ there exists a positive constant $\bar\varepsilon_i$ such that  $\norm{\dot \varepsilon_i}\le \bar\varepsilon_i$.\\
On using (\ref{ldot7}) in (\ref{delta}), the time derivative  of  auxiliary variable  can be written as
% \begin{align}\label{delta1}
%     \dot \delta_i&=\dot \varepsilon_i-k_{\varepsilon_i}\left(Q_i W_i^{*T}\phi_i(\bar z_i)+\beta_i z_{i+1}+\beta_i\vartheta_{i}\right.\notag\\
%     &\quad\left.-Q_i\frac{z_i}{\psi_i}\frac{\partial \psi_i}{\partial t}+\varepsilon_i\right).
% \end{align}
\begin{align}\label{delta1}
    \dot \delta_i&=\dot \varepsilon_i-k_{\varepsilon_i}\left( W_i^{*T}\phi_i(\bar z_i)+\beta_i x_{i+1}\right.\notag\\
    &\quad\left.-Q_i\frac{z_i}{\psi_i}\frac{\partial \psi_i}{\partial t}+\varepsilon_i\right).
\end{align}
For the estimation of the auxiliary variable, its observer dynamics is proposed as 
% \begin{align}\label{deltad}
%     \dot {\hat \delta}_i=-k_{\varepsilon_i}(Q_i\hat W_i^{T}\phi_i(\bar z_i)+\beta_iz_{i+1}+\beta_i\vartheta_{i}-Q_i\frac{z_i}{\psi_i}\frac{\partial \psi_i}{\partial t}+\hat \varepsilon_i).
% \end{align}
\begin{align}\label{deltad}
    \dot {\hat \delta}_i=-k_{\varepsilon_i}(\hat W_i^{T}\phi_i(\bar z_i)+\beta_ix_{i+1}-Q_i\frac{z_i}{\psi_i}\frac{\partial \psi_i}{\partial t}+\hat \varepsilon_i).
\end{align}
Using (\ref{delta}), the estimate of observer variable $\varepsilon_i$ can be obtained as
\begin{align}\label{est}
    \hat \varepsilon_i=\hat \delta_i+k_{\varepsilon_i}L_i.
\end{align}
Using (\ref{delta}) and (\ref{est}), the estimation error of the auxiliary variable can be written as 
\begin{align}\label{obvb}
   \Tilde{\delta}_i=\delta_i-\hat\delta_i=\varepsilon_i-\hat\varepsilon_i=\Tilde{\varepsilon}_i
\end{align}
where $\Tilde{\varepsilon}_i$ is an observer variable estimation error. Now, on  subtracting (\ref{deltad}) from  (\ref{delta1}), we get observer error dynamics as
\begin{align}\label{dm}
   \dot {\Tilde{\varepsilon}}_i=\dot {\Tilde{\delta}}_i=\dot\delta_i-\dot{\hat\delta}_i=\dot\varepsilon_i-k_{\varepsilon_i}\left(-\Tilde{W}^T_i\phi_i(\bar z_i)+\Tilde{\varepsilon}_i\right)
\end{align}
where $\Tilde W_i=\hat W_i-W^*_i$ and $k_{\varepsilon_i} >0$ is an observer gain.\\
The stability analysis of the designed disturbance observer is done further by constructing a Lyapunov function composed of observer variable estimation error $\Tilde{\varepsilon}_i$. In the following section, stability analysis along with the controller design have been achieved. 
\vspace{-0.5cm}
\subsection{Controller design and stability analysis}
In this section, based on the decoupled backstepping method \cite{1220768} a robust adaptive controller is designed such that the output of the system (\ref{sys}) tracks its desired output and all the state variable do not contravene their SVICs. The controller is designed in some steps, and these are as follows\\
\textbf{Step 1:} Consider a Lyapunov function candidate $V_1$ which contains a BLF (\ref{L}) as one of the function
\begin{align}\label{v1}
    V_1=L_1+\frac{1}{2}\tilde{\varepsilon}^2_1+\frac{1}{2\lambda_1}\Tilde{W}^T_1\Tilde{W}_1.
\end{align}
The time derivative of (\ref{v1}), gives
\begin{align}\label{v1d}
    \dot V_1=\dot L_1+{\tilde{\varepsilon}}_1\dot {\tilde{\varepsilon}}_1+\frac{1}{\lambda_1}{\Tilde{W}}^T_1\dot{\hat{W}}_1.
\end{align}
On using (\ref{ldot7}) for $i=1$ in (\ref{v1d}) 
\begin{align}\label{ldot80}
    \dot V_1&= W_1^{*T}\phi_1(\bar z_1)+\beta_1x_{2}-Q_1\frac{z_1}{\psi_1}\frac{\partial \psi_1}{\partial t}+\varepsilon_1\notag\\
    &\quad+{\tilde{\varepsilon}}_1\dot {\tilde{\varepsilon}}_1+\frac{1}{\lambda_1}{\Tilde{W}}^T_1\dot{\hat{W}}_1.
\end{align}
On using (\ref{err}) for $i=1$ in (\ref{ldot80}), $\dot V_1$ can be written as 
\begin{align}\label{ldot8}
    \dot V_1&= W_1^{*T}\phi_1(\bar z_1)+\beta_1z_{2}+\beta_1\vartheta_{1}-Q_1\frac{z_1}{\psi_1}\frac{\partial \psi_1}{\partial t}+\varepsilon_1\notag\\
    &\quad+{\tilde{\varepsilon}}_1\dot {\tilde{\varepsilon}}_1+\frac{1}{\lambda_1}{\Tilde{W}}^T_1\dot{\hat{W}}_1.
\end{align}
In (\ref{ldot8}) if   $\vartheta_1$ is designed  as $\vartheta_1=\frac{2A_{1}}{\pi}\arctan \left(\frac{\pi  v_{1}}{2A_{1}}\right)$ where  $v_{1}=v^{(1)}_{1}+v^{(2)}_{1}$ such that,
$v^{(2)}_1=\beta_1(\sigma^{(1)}_1+\rho(\sigma^{(2)}_1-\sigma^{(1)}_1))$ then based on Lemma 2, we can say  $\beta_1\vartheta_{1}\le\beta_1 v^{(1)}_{1}$. \\
\textit{Note:} In order to  apply Lemma 2, consider the variable $v^{(1)}$, $v^{(2)}$, $\sigma^{(1)}_1$, $\sigma^{(1)}_2$, and $\beta_1$  as  $v_1$, $v_2$, $\sigma_1$, $\sigma_2$, and $\mathfrak{h}$ of Lemma 2,    respectively.\\
Whence, we can write  (\ref{ldot8}) as
\begin{align}\label{ldot9}
    \dot V_1&\le W_1^{*T}\phi_1(\bar z_1)+\beta_1z_{2}+\beta_1 v^{(1)}_{1}-Q_1\frac{z_1}{\psi_1}\frac{\partial \psi_1}{\partial t}+\varepsilon_1\notag\\
    &\quad+{\tilde{\varepsilon}}_1\dot {\tilde{\varepsilon}}_1+\frac{1}{\lambda_1}{\Tilde{W}}^T_1\dot{\hat{W}}_1.
\end{align}

Using the Nussbaum gain function, $v^{(1)}_1$ is designed  as
\begin{align}
    v^{(1)}_1&=\mathcal{N}_1(\zeta_1)\alpha_1,~\text{where} \label{vv1}\\
    \dot \zeta_1&=\alpha_1,~\text{and}\label{g1}\\
    \alpha_1&=k_1Q_1z_1+\hat W_1^T\phi_1(\bar z_1)+\hat\varepsilon_1+\frac{1}{8}k_{\varepsilon_1}^4\notag\\
    &\quad+\frac{3}{4}-Q_1\frac{z_1}{\psi_1}\frac{\partial \psi_1}{\partial t},~ (k_1>0 \text{~is a controller gain})\label{a1}
\end{align}
and an adaptive law for $\hat {W}_1$ is  designed as
\begin{align}\label{upd}
    \dot{\hat W}_1=\lambda_1\left(\phi_1(\bar z_1)-k_{\varepsilon_1}^2\hat W_1 -\eta_1\hat W_1\right).
\end{align}
On using (\ref{vv1}) and (\ref{g1}) in (\ref{ldot9}), $\dot V_1$  can be written as
\begin{align}\label{ldott10}
    \dot V_1&\le W_1^{*T}\phi_1(\bar z_1)+\beta_1z_{2}+\beta_1 \mathcal{N}_1(\zeta_1)\dot \zeta_1+\dot \zeta_1-\alpha_1\notag\\
    &\quad-Q_1\frac{z_1}{\psi_1}\frac{\partial \psi_1}{\partial t}+\varepsilon_1+{\tilde{\varepsilon}}_1\dot {\tilde{\varepsilon}}_1+\frac{1}{\lambda_1}{\Tilde{W}}^T_1\dot{\hat{W}}_1.
\end{align}
On substituting $\alpha_1$ from (\ref{a1}), i.e.  in (\ref{ldott10})
\begin{align}\label{ldottt10}
    \dot V_1&\le- k_1Q_1z_1 - \tilde W_1^{T}\phi_1(\bar z_1)+\beta_1z_{2}+\beta_1 N_1(\zeta_1)\dot \zeta_1+\dot \zeta_1\notag\\
    &\quad +\tilde \varepsilon_1+{\tilde{\varepsilon}}_1\dot {\tilde{\varepsilon}}_1+\frac{1}{\lambda_1}{\Tilde{W}}^T_1\dot{\hat{W}}_1-\frac{1}{8}k_{\varepsilon_1}^4 -\frac{3}{4}.
\end{align}
  Following are some inequalities which will be used in each step of the controller design with change in lower indices.
  \begin{enumerate}[i)]
\item First inequality  is for  the first term of (\ref{ldottt10}), i.e.  $k_1Q_1z_1$.\\
Using $Q_1=\frac{z_1}{\psi_1^2-z_1^2}$ as given in (\ref{Qi}) , we have
\begin{align}
    Q_1z_1=\frac{z_1^2}{\psi_1^2-z_1^2} \label{qz}.
\end{align}
On using the inequality given in Lemma 3,  (\ref{qz}) can be written as 
\begin{align}
    - \frac{1}{2}Q_1z_1=-\frac{1}{2}\frac{z_1^2}{\psi_1^2-z_1^2}\le- \frac{1}{2}\log \frac{\psi_1^2}{\psi_1^2-z_1^2} \label{inl}.
\end{align}
 Using (\ref{L}), the above inequality (\ref{inl}) can be written as
 \begin{align}
     - k_1Q_1z_1\le-2k_1 L_1 \label{Fi}.
 \end{align}
\end{enumerate}
\begin{enumerate}[i)]
\setcounter{enumi}{1}
    \item Second inequality is for  the third term of (\ref{ldottt10}), i.e. $\beta_1 z_2$. \begin{align}\label{Si}
        \beta_1 z_2\le\frac{1}{4}+\beta_1^2 z_2^2.
    \end{align}
\end{enumerate}
 \begin{enumerate}[i)]
\setcounter{enumi}{2}
\item Third inequality is for  the sixth and seventh term of (\ref{ldottt10}), i.e. $\left(\tilde \varepsilon_1+{\tilde{\varepsilon}}_1\dot {\tilde{\varepsilon}}_1\right)$.
\end{enumerate}
Following (\ref{dm}) for $i=1$,  $\tilde \varepsilon_1+{\tilde{\varepsilon}}_1\dot {\tilde{\varepsilon}}_1$  can be written as
\begin{align}\label{in1}
    \tilde \varepsilon_1+{\tilde{\varepsilon}}_1\dot {\tilde{\varepsilon}}_1&=\tilde \varepsilon_1+{\tilde{\varepsilon}}_1\dot\varepsilon_1-{\tilde{\varepsilon}}_1 k_{\varepsilon_1}\left(-\Tilde{W}^T_1\phi_1(\bar z_1)+\Tilde{\varepsilon}_1\right),\notag\\
    &=\tilde \varepsilon_1+{\tilde{\varepsilon}}_1\dot\varepsilon_1+{\tilde{\varepsilon}}_1 k_{\varepsilon_1}\Tilde{W}^T_1\phi_1(\bar z_1)-k_{\varepsilon_1}{\tilde{\varepsilon}}^2_1.
\end{align}
Following (\ref{var}), Assumption 1 and 2, we have  $\dot \varepsilon_i$ is bounded. Let say there exists a positive constant $\bar\varepsilon_i$  such that $\forall i \in \mathbb{N}_n$
\begin{align}\label{new}
    \abs{\dot \varepsilon_i}\le \bar\varepsilon_i
\end{align}
Applying Young's inequality in (\ref{in1}), and following (\ref{bnd1}) and (\ref{new}), we have
\begin{align}\label{Ti}
    \tilde \varepsilon_1+{\tilde{\varepsilon}}_1\dot {\tilde{\varepsilon}}_1 &\le\frac{\tilde \varepsilon_1^2}{2}+\frac{1}{2}+\frac{\tilde \varepsilon_1^2}{2}+\frac{{\bar \varepsilon}_1^2}{2}+\frac{{\tilde{\varepsilon}}_1^2\bar{\phi}_1^2}{2}+\frac{1}{2}k_{\varepsilon_1}^2\norm{\Tilde{W}_1}^2\notag\\
    &\quad-k_{\varepsilon_1}{\tilde{\varepsilon}}^2_1,\notag\\
     &=-\tilde \varepsilon_1^2\left(k_{\varepsilon_1}-1-\frac{\bar{\phi}_1^2}{2}\right)+\frac{1}{2}+\frac{{\bar \varepsilon}_1^2}{2}+\frac{1}{2}k_{\varepsilon_1}^2\norm{\Tilde{W}_1}^2.
\end{align}
\begin{enumerate}[i)]
 \setcounter{enumi}{3}
    \item Fourth inequality is for  the eighth term of (\ref{ldottt10}), i.e. $\frac{1}{\lambda_1}{\Tilde{W}}^T_1\dot{\hat{W}}_1$.
\end{enumerate}
Simplifying the expression $\frac{1}{\lambda_1}{\Tilde{W}}^T_1\dot{\hat{W}}_1$ using (\ref{upd}), we have
\begin{align}\label{win}
    \frac{1}{\lambda_1}{\Tilde{W}}^T_1\dot{\hat{W}}_1&=\Tilde{W}^T_1\phi_1(\bar z_1)-k_{\varepsilon_1}^2\Tilde{W}^T_1\hat W_1-\eta_1\Tilde{W}^T_1\hat W_1
\end{align}
Using the inequality below
\begin{align}
    -{\tilde W_1^T}{\hat{W}}_1\le\frac{1}{2}\left(\norm{W_1^*}^2-\norm{\tilde W_1}^2\right)
\end{align}
in (\ref{win}), we have
\begin{align}\label{win1}
 \frac{1}{\lambda_1}{\Tilde{W}}^T_1\dot{\hat{W}}_1&\le \Tilde{W}^T_1\phi_1(\bar z_1)+\frac{1}{2}k_{\varepsilon_1}^2\norm{W_1^*}^2-\frac{1}{2}k_{\varepsilon_1}^2\norm{\tilde W_1}^2\notag\\
    &\quad+\frac{\eta_1}{2}\norm{W_1^*}^2-\frac{\eta_1}{2}\norm{\tilde W_1}^2.
\end{align}
On applying Young's inequality in the second term of  (\ref{win1}), we have
\begin{align}\label{Foi}
    \frac{1}{\lambda_1}{\Tilde{W}}^T_1\dot{\hat{W}}_1&\le \Tilde{W}^T_1\phi_1(\bar z_1)+\frac{1}{8}k_{\varepsilon_1}^4+\frac{1}{2}\norm{W_1^*}^4\notag\\
    &\quad-\frac{1}{2}k_{\varepsilon_1}^2\norm{\tilde W_1}^2+\frac{\eta_1}{2}\norm{W_1^*}^2-\frac{\eta_1}{2}\norm{\tilde W_1}^2.
\end{align}  
Using all the four inequalities (\ref{Fi}), (\ref{Si}), (\ref{Ti}), and (\ref{Foi}) in  (\ref{ldottt10}), we have 
\begin{align}\label{ldot11}
    \dot V_1&\le  \beta_1 \mathcal{N}_1(\zeta_1)\dot \zeta_1+\dot \zeta_1 -2k_1L_1 +\beta_1^2z_{2}^2\notag\\
    &\quad -\tilde \varepsilon_1^2\left(k_{\varepsilon_1}-1-\frac{\bar{\phi}_1^2}{2}\right)-\frac{\eta_1}{2}\norm{\tilde W_1}^2+\varrho_1,
\end{align}
where $\varrho_1=\frac{{\bar \varepsilon}_1^2}{2}+\frac{1}{2}\norm{W_1^*}^4+\frac{\eta_1}{2}\norm{W_1^*}^2.$\\
The equation (\ref{ldot11}) can be further written as
\begin{align}\label{ldot12}
    \dot V_1&\le-\mu_1V_1+  \beta_1 \mathcal{N}_1(\zeta_1)\dot \zeta_1+\dot \zeta_1 +\beta_1^2z_{2}^2+\varrho_1,
\end{align}
where  $\mu_1=\min\left(2k_1,2\left(k_{\varepsilon_1}-1-\frac{\bar{\phi}_1^2}{2}\right),\lambda_1\eta_1\right)$.\\

\textit{Remark 2:}  In the  decoupled backstepping design, we will  seek for the boundedness of $z_2$ in  the next step of the  design rather than cancellation of  $\beta_1^2z_2^2$. $\hfill \square$\\
 On multiplying both sides of (\ref{ldot12}) by $e^{\mu_1 t}$, we have
\begin{align}\label{ldot13}
\frac{d(V_1(t)e^{\mu_1 t})}{dt}\le\left(\beta_i{\mathcal{N}}_1(\zeta_1)\dot\zeta_1+\dot\zeta_1+\beta_1^2z_{2}^2+\varrho_1\right)e^{\mu_1 t}. 
\end{align}
On integrating (\ref{ldot13}) over $\left[0,t\right]$, gives
\begin{align}
 e^{\mu_1 t}V_1(t)&\le V_1(0) + \int_{0}^{t}{ \left(\beta_i{\mathcal{N}}_1(\zeta_1)+1\right) \dot\zeta_1e^{\mu_1\tau}d\tau}\notag\\
 &\quad +\beta_1^2\int_{0}^{t}{z_{2}^2e^{\mu_1 \tau}d\tau}+\frac{\varrho_1 e^{\mu_1 t}}{\mu_1}-\frac{\varrho_1}{\mu_1} \label{ldot14}.
\end{align}
On multiplying both sides of (\ref{ldot14}) by $e^{-\mu_1 t}$, we have
\begin{align}
 V_1(t)&\le e^{-\mu_1 t}V_1(0) + e^{-\mu_1 t}\int_{0}^{t}{ \left(\beta_i{\mathcal{N}}_1(\zeta_1)+1\right) \dot\zeta_1e^{\mu_1\tau}d\tau}\notag\\
 &\quad +e^{-\mu_1 t}\beta_1^2\int_{0}^{t}{z_{2}^2e^{\mu_1 \tau}d\tau}+\frac{\varrho_1 }{\mu_1}-\frac{\varrho_1e^{-\mu_1 t}}{\mu_1}. \label{ldot15}
\end{align}
 Since, $0<e^{-\mu_1 t}\le1$, we can write (\ref{ldot15}) as
 \begin{align}
 V_1(t)&\le V_1(0) + e^{-\mu_1 t}\int_{0}^{t}{ \left(\beta_1{\mathcal{N}}_1(\zeta_1)+1\right) \dot\zeta_1e^{\mu_1\tau}d\tau}\notag\\
 &\quad +e^{-\mu_1 t}\beta_1^2\int_{0}^{t}{z_{2}^2e^{\mu_1 \tau}d\tau}+\frac{\varrho_1 }{\mu_1}-\frac{\varrho_1e^{-\mu_1 t}}{\mu_1}. \label{ldot166}
\end{align}
We can rewrite (\ref{ldot166}) as
 \begin{align}
 V_1(t)&\le V_1(0) + e^{-\mu_1 t}\int_{0}^{t}{ \left(\beta_1{\mathcal{N}}_1(\zeta_1)+1\right) \dot\zeta_1e^{\mu_1\tau}d\tau}\notag\\
 &\quad +e^{-\mu_1 t}\beta_1^2\int_{0}^{t}{z_{2}^2e^{\mu_1 \tau}d\tau}+\frac{\varrho_1 }{\mu_1}. \label{ldot1666}
\end{align}
\textit{Remark 3:} In (\ref{ldot166}), if there would have been no extra term, i.e.  $e^{-\mu_1 t}\beta_1^2\int_{0}^{t}{z_{2}^2e^{\mu_1 \tau}d\tau}$, then using Lemma 1, we may have shown that $V_1(t),\zeta_1$ and $z_1,\hat W_1, \hat \varepsilon_1$ are all uniformly ultimately bounded. However, if we can show  $z_2$ is bounded, then using the following relation 
\begin{align}
    e^{-\mu_1 t}\int_{0}^{t}{\beta_1^2z_{2}^2e^{\mu_1 \tau}d\tau}&\le e^{-\mu_1 t}\beta_1^2\underset{\tau\in[0,t]}{\sup} z_{2}^2\int_{0}^{t}{e^{\mu_1 \tau}d\tau} \notag\\
    &\le \frac{\beta_1^2\underset{\tau\in[0,t]}{\sup} z_{2}^2}{\mu_1}, \label{bound1}
\end{align}
we can say that $e^{-\mu_1 t}\beta_1^2\int_{0}^{t}{z_{2}^2e^{\mu_1 \tau}d\tau}$ is bounded. Consequently using Lemma 1, we will be able to show $V_1(t),\zeta_1$ and $z_1,\hat W_1, \hat \varepsilon_1$ are also bounded. Again to show $z_2$ is bounded, we need to follow similar steps. The process will be recursive until we do not  have   $\beta_i^2z_{i+1}^2$ in the derivative of Lyapunov function. $\hfill \square$

\noindent\textbf{Step i} \textit{$(2\le i \le n-1)$:} Consider a Lyapunov function candidate $V_i$ which has $L_i$ as one of its component
\begin{align}\label{vi}
    V_i=L_i+\frac{1}{2}\tilde{\varepsilon}^2_i+\frac{1}{2\lambda_i}\Tilde{W}^T_i\Tilde{W}_i.
\end{align}
On taking the time derivative of (\ref{vi}) and using (\ref{ldot7}), we have
\begin{align}\label{vid}
    \dot V_i&=W_i^{*T}\phi_i(\bar z_i)+\beta_ix_{i+1}-Q_i\frac{z_i}{\psi_i}\frac{\partial \psi_i}{\partial t}\notag\\
    &\quad+\varepsilon_i+{\tilde{\varepsilon}}_i\dot {\tilde{\varepsilon}}_i+\frac{1}{\lambda_i}{\Tilde{W}}^T_i\dot{\hat{W}}_i.
\end{align}
On using (\ref{err}) in (\ref{vid})
\begin{align}\label{vid1}
    \dot V_i&=W_i^{*T}\phi_i(\bar z_i)+\beta_i z_{i+1}+\beta_i\vartheta_{i}-Q_i\frac{z_i}{\psi_i}\frac{\partial \psi_i}{\partial t}\notag\\
    &\quad+\varepsilon_i+{\tilde{\varepsilon}}_i\dot {\tilde{\varepsilon}}_i+\frac{1}{\lambda_i}{\Tilde{W}}^T_i\dot{\hat{W}}_i.
\end{align}
In (\ref{vid1}), if   $\vartheta_i$ is designed  as $\vartheta_i=\frac{2A_{i}}{\pi}\arctan \left(\frac{\pi  v_{i}}{2A_{i}}\right)$ where  $v_{i}=v^{(1)}_{i}+v^{(2)}_{i}$ such that,
$v^{(2)}_i=\beta_i(\sigma^{(1)}_i+\rho(\sigma^{(2)}_i-\sigma^{(1)}_i))$ then based on Lemma 2, we can say  $\beta_i\vartheta_{i}\le\beta_i v^{(1)}_{i}$. Consequently, we can write  (\ref{vid1}) as
\begin{align}\label{vid2}
    \dot V_i&\le W_i^{*T}\phi_i(\bar z_i)+\beta_i z_{i+1}+\beta_iv_{i}^{(1)}-Q_i\frac{z_i}{\psi_i}\frac{\partial \psi_i}{\partial t}\notag\\
    &\quad+\varepsilon_i+{\tilde{\varepsilon}}_i\dot {\tilde{\varepsilon}}_i+\frac{1}{\lambda_i}{\Tilde{W}}^T_i\dot{\hat{W}}_i.
\end{align}
Designing $\vartheta_i$   and adaptive law as 
\begin{align}
    v_i&=v_i^{(1)}+v_i^{(2)},~ \text{where}\\
    v^{(1)}_i&=\mathcal{N}_i(\zeta_i)\alpha_i, \label{vvi}\\
    \dot \zeta_i&=\alpha_i,\label{gi}\\
    \alpha_i&=k_iQ_iz_i+\hat W_i^T\phi_i(\bar z_i)+\hat\varepsilon_i+\frac{1}{8}k_{\varepsilon_i}^4\notag\\
    &\quad+\frac{3}{4}-Q_i\frac{z_i}{\psi_i}\frac{\partial \psi_i}{\partial t}, ~\text{and} \label{ai}\\
    \dot{\hat W}_i&=\lambda_i\left(\phi_i(\bar z_i)-k_{\varepsilon_i}^2\hat W_i -\eta_i\hat W_i\right). \label{updi}
\end{align}
Following the same procedure as step 1, we have
\begin{align}
 V_i(t)&\le V_i(0) + e^{-\mu_i t}\int_{0}^{t}{ \left(\beta_i{\mathcal{N}}_i(\zeta_i)+1\right) \dot\zeta_ie^{\mu_i\tau}d\tau}\notag\\
 &\quad +e^{-\mu_i t}\beta_i^2\int_{0}^{t}{z_{i+1}^2e^{\mu_i \tau}d\tau}+\frac{\varrho_i }{\mu_i} \label{ldot16}
\end{align}
where $\mu_i=\min\left(2k_i,2\left(k_{\varepsilon_i}-1-\frac{\bar{\phi}_i^2}{2}\right),\lambda_i\eta_i\right)$, $\varrho_i=\frac{{\bar \varepsilon}_i^2}{2}+\frac{1}{2}\norm{W_i^*}^4+\frac{\eta_i}{2}\norm{W_i^*}^2$ and \begin{align}
    e^{-\mu_i t}\beta_i^2\int_{0}^{t}{z_{i+1}^2e^{\mu_i \tau}d\tau}&\le e^{-\mu_i t}\beta_i^2\underset{\tau\in[0,t]}{\sup} z_{i+1}^2\int_{0}^{t}{e^{\mu_i \tau}d\tau} \notag\\
    &\le \frac{\beta_i^2\underset{\tau\in[0,t]}{\sup} z_{i+1}^2}{\mu_i}. \label{boundi}
\end{align}
\textit{Remark 4:} Similar to previous discussion in Remark 3, we can apply Lemma 1 to show $V_i(t),\zeta_i$ and $z_i,\hat W_i, \hat \varepsilon_i$ are all uniformly ultimately bounded, provided $z_{i+1}$ is bounded. $\hfill \square$\\
\noindent \textbf{Step n:} 
Similar to previous step, consider a Lyapunov function candidate $V_n$ which contains a BLF (\ref{L}) 
\begin{align}\label{vn}
    V_n=L_n+\frac{1}{2}\tilde{\varepsilon}^2_n+\frac{1}{2\lambda_n}\Tilde{W}^T_n\Tilde{W}_n.
\end{align}
On taking the time derivative of (\ref{vn}) and using (\ref{ldot7}) with $x_{n+1}=u$, we have
\begin{align}\label{vnd}
    \dot V_n&=W_n^{*T}\phi_i(\bar z_n)+\beta_nu-Q_n\frac{z_n}{\psi_n}\frac{\partial \psi_n}{\partial t}\notag\\
    &\quad+\varepsilon_n+{\tilde{\varepsilon}}_n\dot {\tilde{\varepsilon}}_n+\frac{1}{\lambda_n}{\Tilde{W}}^T_n\dot{\hat{W}}_n.
\end{align}
\textit{Remark 5:}It is to be noted that unlike the previous steps, where  we have replaced $x_{i+1}$ with $z_{i+1}+\vartheta_i$ using the relation (\ref{err}), here control input is directly available for design. We can also observe that as compared  to (\ref{ldot8}) and (\ref{vid1}), (\ref{vnd}) doesn't involve extra term $\beta_i z_{i+1}$. $\hfill \square$\\
 \textit{Remark 6:} Since (\ref{vnd}) doesn't involve additional term $\beta_n z_{n+1}$, we can't apply inequality similar to the second inequality of step 1, so there will be no extra term ${1}/{4}+\beta _n^2 z_{n+1}^2$. The effect  of this can be seen in the following design procedure. $\hfill \square$\\  
 The control input and adaptive law are designed as the previous step 
 \begin{align}
    u&=N_n(\zeta_n)\alpha_n, \label{vvn}\\
    \dot \zeta_n&=\alpha_n,\label{gn}\\
    \alpha_n&=k_nQ_nz_n+\hat W_n^T\phi_n(\bar z_n)+\hat\varepsilon_n+\frac{1}{8}k_{\varepsilon_n}^4\notag\\
    &\quad+\frac{1}{2}-Q_n\frac{z_n}{\psi_n}\frac{\partial \psi_n}{\partial t}, ~ \text{and}\label{an}\\
    \dot{\hat W}_n&=\lambda_n\left(\phi_n(\bar z_n)-k_{\varepsilon_n}^2\hat W_n -\eta_n\hat W_n\right). \label{wn}
 \end{align}
\textit{Remark 7:} It can be seen that as compared  to (\ref{a1}) and (\ref{ai}) in (\ref{an}), $\alpha_n$  has  the term ${1}/{2} (={3}/{4}-{1}/{4})$ in place of ${3}/{4}$. $\hfill \square$\\
 On using (\ref{vvn})-(\ref{wn}) in (\ref{vnd}) and following the same procedure as in the previous steps, we have
 \begin{align}\label{vnd1}
     V_n(t)&\le V_n(0) + e^{-\mu_n t}\int_{0}^{t}{ \left(\beta_n{\mathcal{N}}_n(\zeta_n)+1\right) \dot\zeta_ne^{\mu_n\tau}d\tau}  +\frac{\varrho_n }{\mu_n} , 
 \end{align}
 where $\mu_n=\min\left(2k_n,2\left(k_{\varepsilon_n}-1-\frac{\bar{\phi}_n^2}{2}\right),\lambda_n\eta_n\right) \text{and} ~ \varrho_n=\frac{{\bar \varepsilon}_n^2}{2}+\frac{1}{2}\norm{W_n^*}^4+\frac{\eta_n}{2}\norm{W_n^*}^2.$\\
 In  (\ref{vnd1}),   $V_n(0)+{\varrho_n }/{\mu_n}$ is a constant. Let $c_n=V_n(0)+{\varrho_n }/{\mu_n}$, then using Lemma 1 in (\ref{vnd1}) we can say $V_n(t),\zeta_n$ and $z_n,\hat W_n, \hat \varepsilon_n$ are  uniformly ultimately bounded.
Due to the boundedness of   $z_n$, for $i=n-1$ in (\ref{boundi}) we can say, the integral term $e^{-\mu_{n-1} t}\beta_{n-1}^2\int_{0}^{t}{z_{n}^2e^{\mu_{n-1} \tau}d\tau}$ is bounded. Thus, based on  Lemma 1 and  (\ref{ldot16}) for $i=n-1$ we can conclude that   $V_{n-1}(t),\zeta_{n-1}$ and $z_{n-1},\hat W_{n-1}, \hat \varepsilon_{n-1}$ are  also uniformly ultimately bounded. Similarly, we can prove in that $V_i(t),\zeta_i$ and $z_i,\hat W_i, \hat \varepsilon_i$ are  uniformly ultimately bounded $\forall i \in \mathbb{N}_{n-2}$.
\vspace{-0.2cm}
\section{Boundedness and Convergence}
\textbf{{Theorem 1:}} For a class of system (\ref{sys}), under Assumptions 1-3
and initial error condition $\abs{z_i(0)}<\abs{\psi(\bar x_i(0),0)}$, if the adaptive controller  is designed and  controller parameters are updated as given in Table \ref{tabcon} and Table \ref{tabupdate}, respectively, then the closed-loop system holds the following properties:
\begin{enumerate}[i)]
    \item All the closed-loop signals are uniformly ultimately bounded.
    \item All the states of the system will never contravene their respective SVICs, i.e. $\abs{x_i}<\Psi_i(\bar x_i,t)$.
    \item The closed-loop error signal $z_1$ will converge to a small neighbourhood of zero.
\end{enumerate}
\textbf{{Proof i).}} Following all the steps $1$ to $n$ of controller design and stability analysis, it can be easily proved that all the closed-loop signals are bounded.\\
\textbf{{Proof ii).}} To prove this, we will use proof by contradiction. Let us assume that, for $i=1$ there exists some $t=\mathbb{T}$,  such that $\abs{z_1(\mathbb{T})}$ grows to $\psi(\bar x_{1}(\mathbb{T}),\mathbb{T})$.Then, substituting $\abs{z_{1}(\mathbb{T})}=\psi_1(\bar x_{1}(\mathbb{T}),\mathbb{T})$ in (\ref{L}) makes  $L_1=\frac{1}{2}\log\frac{\psi_1^2}{\psi_1^2-z_1^2}$ unbounded and based on  (\ref{v1}),  $V_1$ involve $L_1$, i.e. $V_1$ will becomes unbounded,  contradicting the previous proved results. Thus, for any $t$, $\abs{z_1(t)}<\psi_1(\bar x_{1}(t),t)$. Similarly, we can prove this  $\forall i \in (\mathbb{N}_{n}-\mathbb{N}_{2})$. Hence, we have  
\begin{align}
\abs{z_i(t)}<\psi_i(\bar x_{i}(t),t), ~~~\forall i\in  \mathbb{N}_n.\label{p1}    
\end{align}
Now, from ($\ref{err}$) we have 
\begin{align}
\abs{x_i}\le \abs{z_i}+\abs{\vartheta_{i-1}}, ~~~\forall i\in  \mathbb{N}_n.\label{p2}    
\end{align}
Using (\ref{nvi}), we can write
\begin{align}
    \abs{\vartheta_{i-1}}\le A_{i-1}.\label{p3}
\end{align}
On using (\ref{p1}) and (\ref{p3}), we can write (\ref{p2}) as
\begin{align}
  \abs{x_i}< \psi(\bar x_{i}(t),t)+A_{i-1}, ~~~\forall i\in  \mathbb{N}_n.\label{p4}   
\end{align}
Rewriting (\ref{p4}), using the relation given in  (\ref{con}), we have
\begin{align}
    \abs{x_i}<\Psi_i(\bar x_i,t), ~~~\forall i\in  \mathbb{N}_n. \label{p5}   
\end{align}
Thus, from  (\ref{p5}) it is proved that all the states of system will never contravene their respective SVICs, i.e. $\abs{x_i}<\Psi(\bar x_i,t)$. \\   
\textbf{{Proof iii).}} Let $C_{\zeta_1}$ be the upper bound of integral term in (\ref{ldot15}) 
\begin{align}\label{boundd}
    e^{-\mu_1 t}&\int_{0}^{t}{ \left(\beta_1{\mathcal{N}}_1(\zeta_1)+1\right) \dot\zeta_1e^{\mu_1\tau}d\tau}\notag\\
    &+e^{-\mu_1 t}\int_{0}^{t}{\beta_1^2z_{2}^2e^{\mu_1 \tau}d\tau}\le C_{\zeta_1}.
\end{align}
 Following (\ref{v1}) and (\ref{L}), and using (\ref{boundd}),  we can write (\ref{ldot15}) as
\begin{align}\label{inep}
    \frac{1}{2}\log\frac{\psi_1^2}{\psi_1^2-z_1^2}\le V_1(t)&\le e^{-\mu_1 t}\left(V_1(0)-\frac{\varrho_1}{\mu_1}\right)+\frac{\varrho_1 }{\mu_1}+C_{\zeta_1}.
\end{align}
On solving the above inequality, we have (\ref{inep}) as
\begin{align}\label{inep1}
    \abs{z_1}\le\psi_1\sqrt{1-e^{-2\frac{\varrho_1 }{\mu_1}-2C_{\zeta_1}}e^{-2\left(V_1(0)-\frac{\varrho_1}{\mu_1}\right)e^{-\mu_1 t}}}
\end{align}
 For $t\rightarrow\infty$ in (\ref{inep1}), we have 
\begin{align}\label{inep2}
     \abs{z_1}\le\psi_1\sqrt{1-e^{-2\frac{\varrho_1 }{\mu_1}-2C_{\zeta_1}}}.
\end{align}
In the above error bound of $z_1$, we can see that $z_1$ can be made arbitrarily small, by selecting the design parameters appropriately.\par
In the next section, to show the effectiveness of the proposed controller, an example has been demonstrated. 
\begin{table}[H]
 \caption{Adaptive controller using NN}
    \centering
    \begin{tabular}{c}
    \toprule
    % \textbf{Control input  and virtual control input}\\
    \midrule
     $\begin{aligned}
     &\quad\quad\quad\quad\quad\quad\quad\quad\quad\textbf{Error variable:}\\
     z_i&=x_i-\vartheta_{i-1}~~~ \forall i \in \mathbb{N}_n. \\
     \textit{Note}:&~\vartheta_0~\text{is a desired output and~} \vartheta_1~\hdots~\vartheta_{n-1}~ \text{are virtual control inputs}.\\ 
     &\quad\quad\quad\quad\quad\quad\quad\quad\textbf{Virtual control input:}\\
     \vartheta_{i}&=\frac{2A_{i}}{\pi}\arctan \left(\frac{\pi  v_{i}}{2A_{i}}\right)~~~ \forall i \in \mathbb{N}_{n-1},~~\text{where}\\
      v_i&=v_i^{(1)}+v_i^{(2)} ~\text{and}\\
    v^{(1)}_i&=N_i(\zeta_i)\alpha_i, ~\text{where}\\
    \alpha_i&=k_iQ_iz_i+\hat W_i^T\phi_i(\bar z_i)+\hat\varepsilon_i+\frac{1}{8}k_{\varepsilon_i}^4+\frac{3}{4}-Q_i\frac{z_i}{\psi_i}\frac{\partial \psi_i}{\partial t},\\
v^{(2)}_i&=\beta_i(\sigma^{(1)}_i+\rho(\sigma^{(2)}_i-\sigma^{(1)}_i)).\\
\textit{Note}:&~ Q_i,~\text{and}~ \sigma^{(1)}_i~\text{and}~\sigma^{(2)}_i \text{can be calculated using~} (\ref{Qi}),\\
&~\text{and}~ \text{ Lemma 2, respectively.} \\
&\quad\quad\quad\quad\quad\quad\quad\quad\quad\textbf{Control input:}\\
u&=N_n(\zeta_n)\alpha_n,~\text{where}\\
\alpha_n&=k_nQ_nz_n+\hat W_n^T\phi_n(\bar z_n)+\hat\varepsilon_n+\frac{1}{8}k_{\varepsilon_n}^4+\frac{1}{2}-Q_n\frac{z_n}{\psi_n}\frac{\partial \psi_n}{\partial t}
\end{aligned}$\\
        \bottomrule
    \end{tabular}
    \label{tabcon}
\end{table}
\begin{table}[H]
\caption{Update laws for the  parameter of controller}
    \centering
    \begin{tabular}{c}
    \toprule
   \textbf{ Update laws}\\
    \midrule
       \textbf{Nussbaum gain:}\\
        $\dot \zeta_i=\alpha_i$\\
        \textbf{NN weight:}\\
        $ \dot{\hat W}_i=\lambda_i\left(\phi_i(\bar z_i)-k_{\varepsilon_i}^2\hat W_i -\eta_i\hat W_i\right) $\\
       \textbf{Disturbance observer:}\\
       $\dot {\hat \delta}_i=-k_{\varepsilon_i}(\hat W_i^{T}\phi_i(\bar z_i)+\beta_ix_{i+1}-Q_i\frac{z_i}{\psi_i}\frac{\partial \psi_i}{\partial t}+\hat \varepsilon_i).$ \\
        \bottomrule
    \end{tabular}
    \label{tabupdate}
\end{table}
 \begin{figure}
    \centering
    \includegraphics[width=7cm,height=5cm,keepaspectratio]{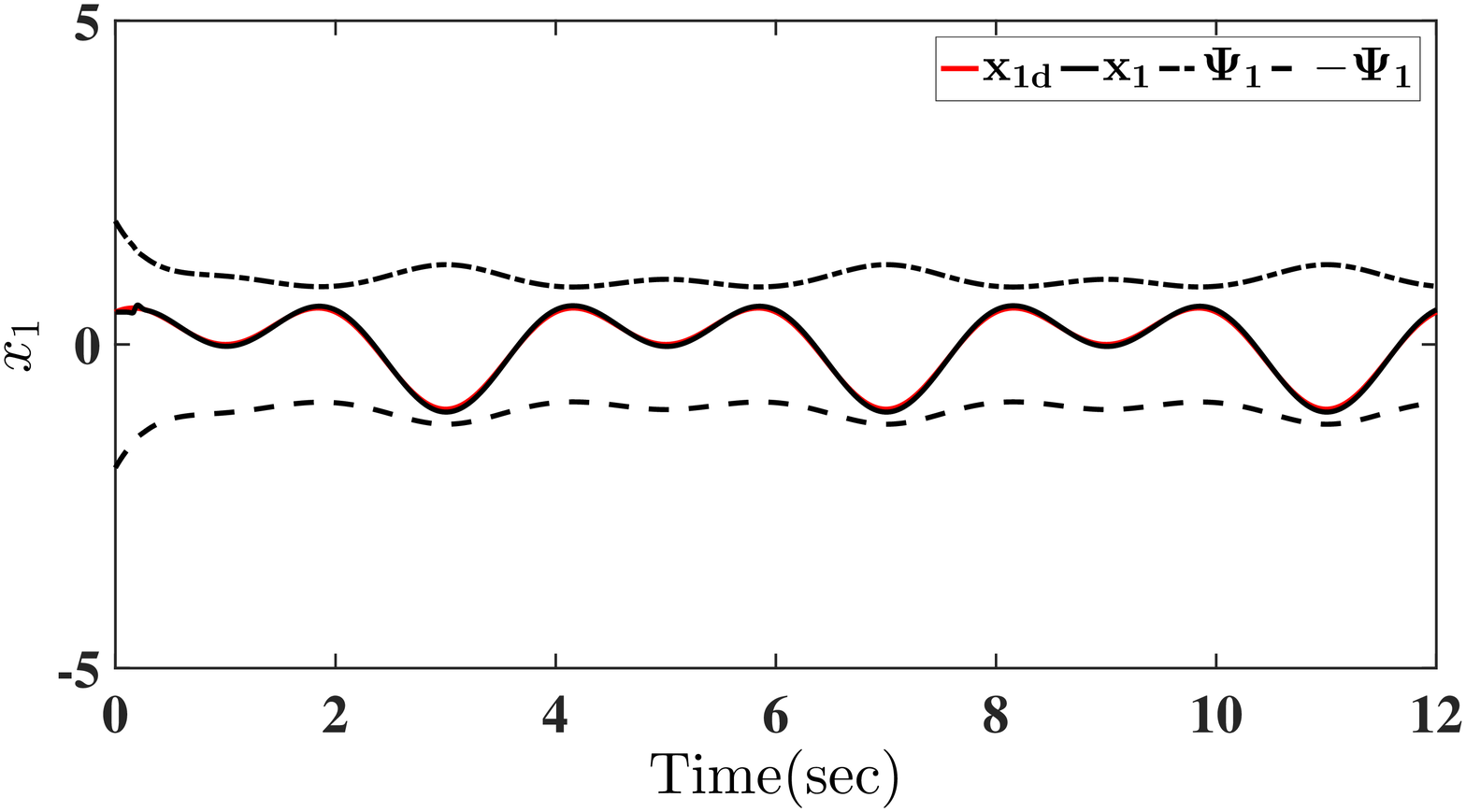}
    \caption{ Output tracking and boundedness performance of  $x_1$.}
    \label{x1}
% \end{figure}
% \begin{figure}
    \centering
    \includegraphics[width=7cm,height=5cm,keepaspectratio]{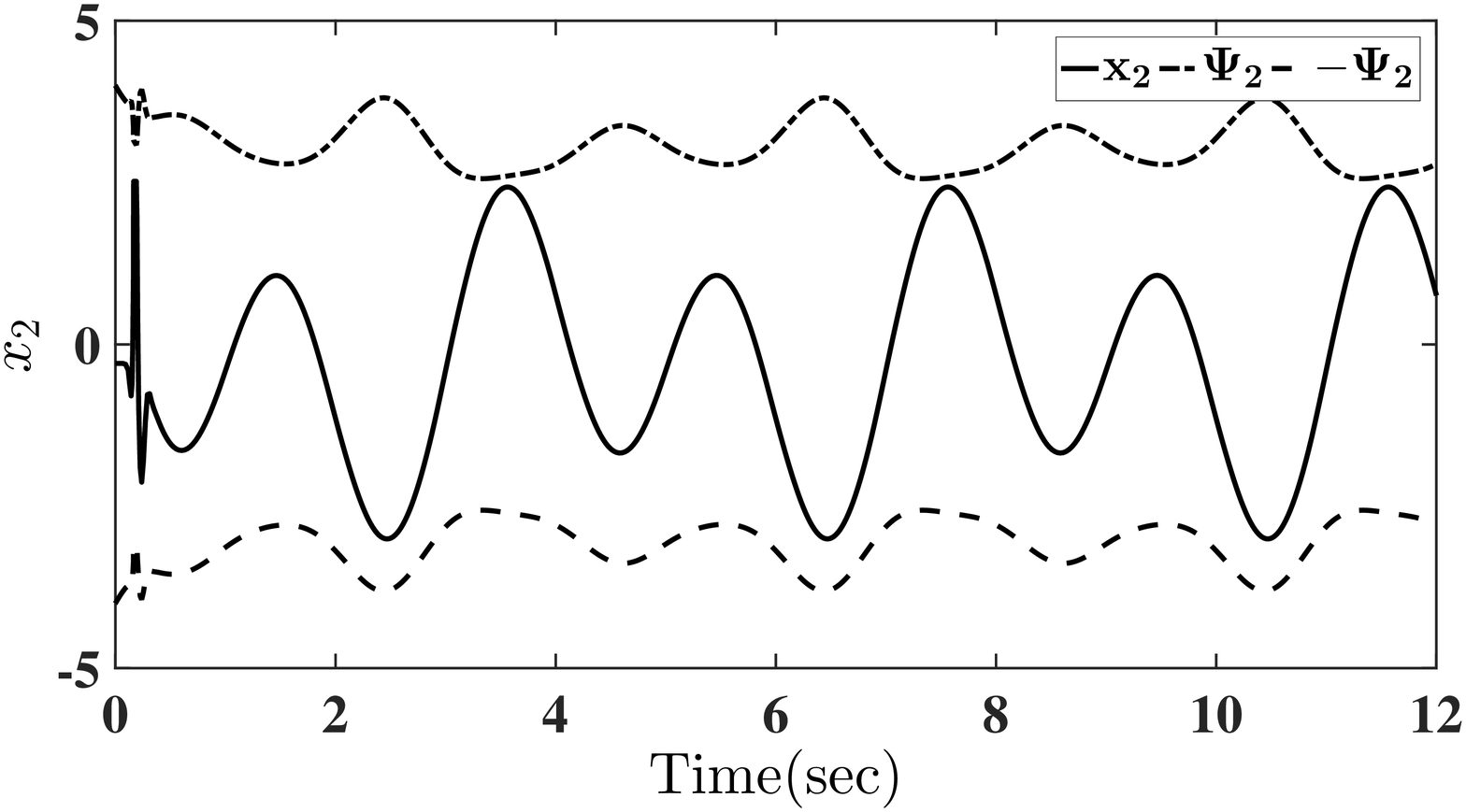}
    \caption{Boundedness performance and trajectory of the state variable $x_2$.}
    \label{x2}
% \end{figure}
% \begin{figure}    
    \centering
    \includegraphics[width=7cm,height=5cm,keepaspectratio]{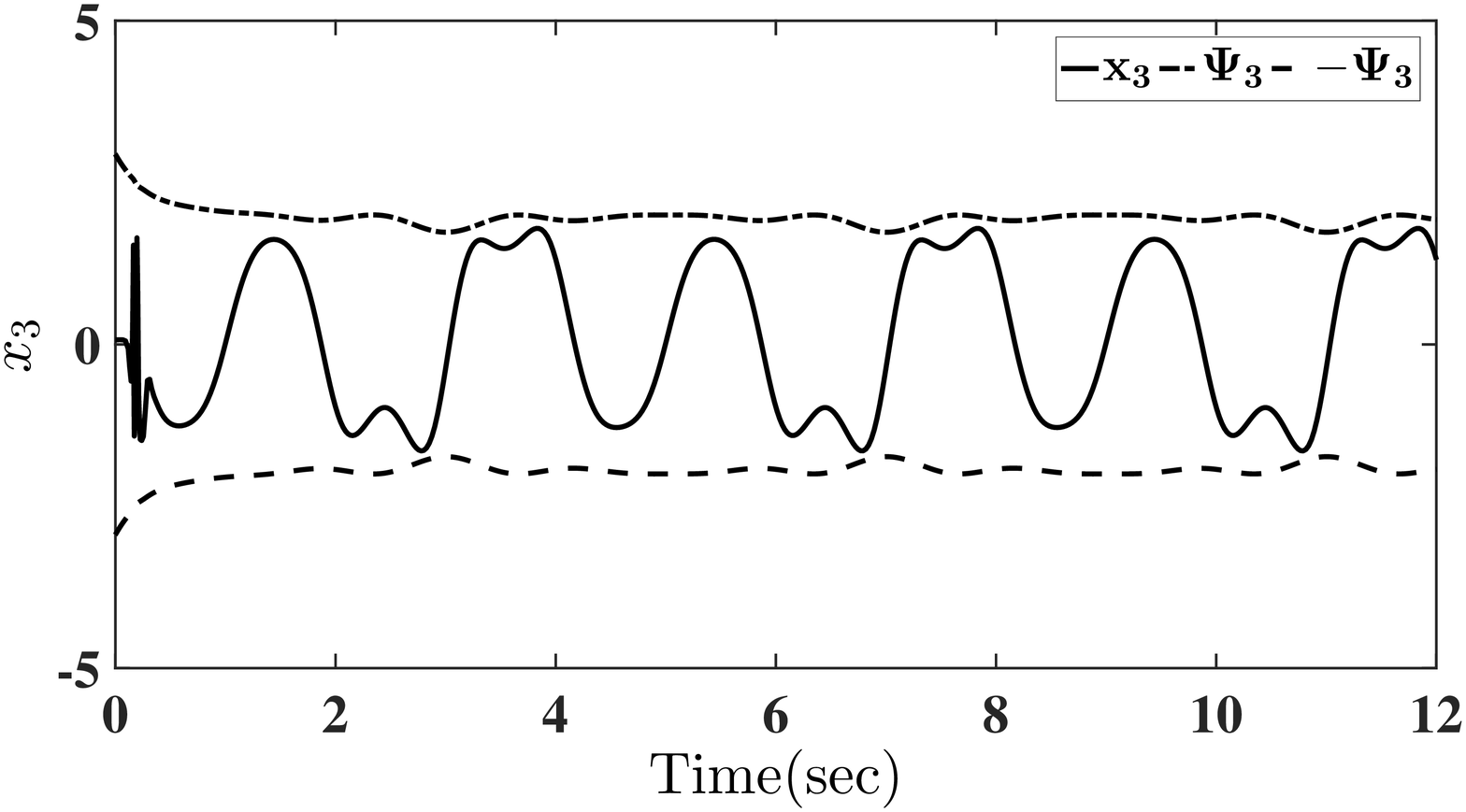}
    \caption{Boundedness performance and trajectory of the state variable $x_3$.}
    \label{x3}
% \end{figure}
% \begin{figure}
\end{figure}
\begin{figure}
    \centering
    \includegraphics[width=7cm,height=5cm,keepaspectratio]{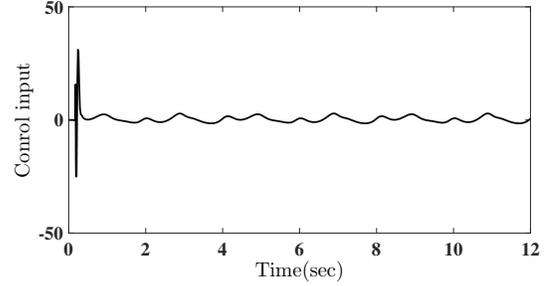}
    \caption{The control input signal.}
    \label{up}
 \end{figure}
\section{Simulation Results and discussion}
Consider a third-order pure feedback nonlinear system \cite{7031439}
\begin{equation}
    \begin{split}
        \dot x_1&=x_2+0.05\cos(x_1)+ d_1\\
\dot x_2&=\frac{1-2^{x_1x_2}}{1+2^{x_1^2}}+x_3+0.1\tanh(x_3)+d_2\\
\dot x_3&=0.2\times 3^{-x_2^2x_3^4}+(0.9+0.05\exp(-x_1^2))u\\
&\quad+0.2\cos(u)+d_3\\
y&=x_1
\end{split}
\end{equation}
 where $x_1,x_2$, and $x_3$ are the states, $u$ is  the control input, and $y=x_1$ is the output of the system. To verify the robustness of  proposed controller, disturbances $d_1=0.2\sin(\pi x_1)$, $d_2=0.2\sin(\pi x_1x_2)$, and $d_3=0.2x_2^2\sin(\pi x_3)$ are considered in  the system.  Let, $y_d=0.5\cos(\pi t)+0.5\sin(0.5\pi t)$ be the system desired output of system, and $\Psi_1=e^{-0.2x_1}+e^{-3t}$, $\Psi_2=e^{-0.2x_2}+e^{-3t}+2\cos(0.5x_1)$, and $\Psi_3=e^{-3t}+2\cos(0.5x_1)$ be the constraints on system states $x_1, x_2$, and $x_3$, respectively. The control objective is to design a control input $u$ such that the system output tracks the desired output $y_d$ and the system states do not contravene their respective constraints, i.e. $\abs{x_1}<\Psi_1$, $\abs{x_2}<\Psi_2$, and $\abs{x_3}<\Psi_3$.\\
 \textit{Remark 8:} It is given in  \cite{4601100} that any SISO system can be expressed in the form of (\ref{sys}).\\
  We have designed an adaptive controller based on Table \ref{tabcon}. Table \ref{tabupdate} has been used to update its parameters. The design parameter and initial values used in the simulation are:
$k_1=k_2=k_3=7$; $k_{\varepsilon_1}=k_{\varepsilon_2}=k_{\varepsilon_3}=6$; $A_0=1$, $A_1=2$, $A_2=2$; $x_1(0)=0.5$, $x_2(0)=-0.3$, $x_3(0)=0$; $\lambda_1=\lambda_2=\lambda_3=10$, and $\zeta_1(0)=0$, $\zeta_2(0)=0$, $\zeta_3(0)=0.2$. The weights of the RBF NN are  chosen as a  $30 \times 1$ dimensional vector, where 30 and 1 represent the number of nodes in the hidden layer and the output of the NN, respectively. 
% Initially each  element of weight matrix is taken as 0.1. The centre  and the width of   for each nodes is chosen as $[0.3~~0.55~~0.80~~1.5]^T$ and $15$, respectively.
\begin{figure}
    \centering
    \includegraphics[width=7cm,height=5cm,keepaspectratio]{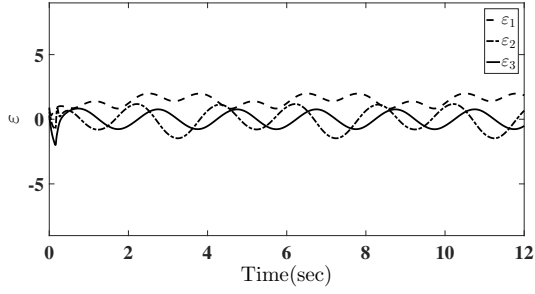}
    \caption{ Trajectory of  $\varepsilon_1$, $\varepsilon_2$, and $\varepsilon_3$.}
    \label{varp}
\end{figure}
\begin{figure}
    \centering
    \includegraphics[width=7cm,height=5cm,keepaspectratio]{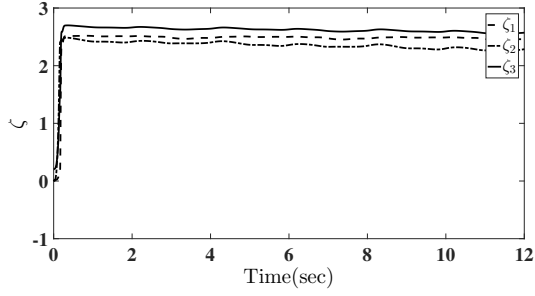}
    \caption{ Trajectories of  $\zeta_1$,$\zeta_2$, and $\zeta_3$.}
    \label{zetap}
\end{figure}
\begin{figure}
    \centering
    \includegraphics[width=7cm,height=5cm,keepaspectratio]{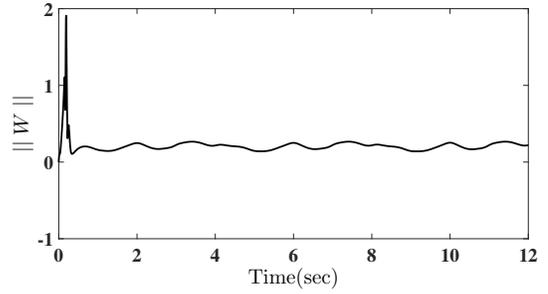}
    \caption{ Norm of NN weights  $W_1$, $W_2$, and $W_3$.}
    \label{nnweight}
\end{figure}\par
 The simulation results are shown in Figs. \ref{x1}-\ref{nnweight}; Figs. \ref{x1}-\ref{x3} show the trajectories of the states and their symmetric time-varying and state-dependent constraints.  From Figs. \ref{x1}-\ref{x3} we can see that all the states are bounded in nature and  doesn't contravene their respective constraints. Also, from Fig. \ref{x1} it can be seen that  the  output tracks its desired trajectory satisfactorily. Furthermore, as proved in Theorem 1, it can be seen from Figs. \ref{x1}-\ref{nnweight} that  all the  signals in the closed-loop system, i.e. control input $u_1$ in Fig. \ref{up},  disturbance observer variables $\varepsilon_1$, $\varepsilon_2$, and $\varepsilon_3$ in Fig. \ref{varp}, Nussbaum gain parameter $\zeta_1, \zeta_2$, and $\zeta_3$ in Fig. \ref{zetap}, and  Norm of NN weights matrix  $W_1$, $W_2$, and $W_3$ in Fig. \ref{nnweight} are bounded in nature. The result thus  shows the effectiveness of the  proposed methodology.
\section{Conclusion}
A robust adaptive backstepping control is proposed for the tracking control of a pure feedback nonlinear system with symmetric SVICs on the state variables. The proposed controller doesn't require prior knowledge of the system dynamics. The neural network is introduced to approximate the behaviour of unknown dynamics which arise during the time derivative of BLF. The use of disturbance observer helped much in making the controller robust and computationally inexpensive by estimating the disturbance along with  NN approximation error and derivative of virtual control input. Through the simulation study, it is shown that all the signals in the closed-loop system are bounded and do not contravene their constraints. In future, this work can be extended for a stochastic pure feedback nonlinear system with asymmetric SVICs on the state variables.
\ifCLASSOPTIONcaptionsoff
 \newpage
\fi
\bibliographystyle{ieeetr}
\bibliography{aa.bib}

% \begin{IEEEbiography}{Pankaj Kumar Mishra}
% received the Master\textquotesingle s degree in control and instrumentation engineering from Motilal Nehru National Institute of Technology (MNNIT)
% Allahabad, India in 2016. 

% His primary research interests include cyber physical systems, smart home automation, and soft computing.
% \end{IEEEbiography}
% \begin{IEEEbiography}{Nishchal Kumar Verma}(SM\textquotesingle13)   
% received Ph.D. degree in electrical engineering from Indian Institute of Technology Delhi, New Delhi, India, in 2007. He is currently an Associate Professor in the department of electrical engineering, Indian Institute of Technology Kanpur, India. His current research interests include the theory and applications of computational intelligence, big data
% analytics, Internet of things, intelligent data mining algorithms, computer vision, and prognosis and health management.

% Dr. Verma is an IETE Fellow. He is currently serving as an Editor
% of the IETE Technical Review journal, an Associate Editor of the IEEE Computational Intelligence Magazine, an Associate Editor of the Transactions of the Institute of Measurement and Control, U.K., and an Editorial Board Member for several reputed journals and conferences.
% \end{IEEEbiography}
\end{document}